\documentclass[12pt]{article}
\usepackage{amsmath,latexsym}
\usepackage{graphicx}
 \def\cen{\centerline}

\begin{document}

\setlength{\unitlength}{1mm}
%\baselineskip 0.1cm
%\large

%\setcounter{page}{1}
%\pagenumbering{arabic}
\title{Global monopole, dark matter and scalar tensor theory}
 \author{F.Rahaman$^*$, M.Kalam$^{\ddag}$,
 R.Mondal $^*$and B.Raychaudhuri $^\dag$}
\date{}
 \maketitle
 \begin{abstract}
                  In this article, we discuss the space-time of a global monopole
                   field as a candidate
                  for galactic dark matter in the context of scalar tensor theory.

  \end{abstract}

 %\bigskip
 %\medskip
  \footnotetext{ Pacs Nos :  04.20 Gz,04.50 + h, 04.20 Jb   \\
 Key words: Global monopole , Scalar tensor theory , Dark matter
\\
 $*$Dept.of Mathematics, Jadavpur University, Kolkata-700 032, India
                                  E-Mail:farook\_rahaman@yahoo.com\\
$\ddag$Dept. of Phys. , Netaji Nagar College for Women, Regent Estate, Kolkata-700092, India.\\
  $\dag$Dept. of Phys. , Surya Sen Mahavidyalaya, Siliguri, West Bengal,  India
}
 %   \mbox{} \hspace{.2in}

    \mbox{} \hspace{.2in}   It is believed that our Universe in the past was denser and
    hotter. So the early stage of the Universe was in a systematic phase
and there were no
    topological defects. As the Universe expands, it cools down from its hot initial
state [1-3]. So the early
     Universe had undergone a number of phase transitions. Phase transitions in the early
     Universe can give rise to various forms of topological defects. They can be monopoles,
     comic strings or domain walls.  Among them monopoles and cosmic string are well
      studied for their cosmological as well as astrophysical implications.
      Monopoles are point like topological defects that may arise during phase transitions
      in the early universe. There has been a fairly large amount of discussions [4-24] on the gravitational
        field of global monopoles beginning with the work of Barriola and Vilenkin (BV) [25]. According to BV,
         global monopoles are the
        configurations with energy density
decreasing with the distance
        as $\frac{1}{r^2}$.
        Recent observations indicate that the  galaxies, cluster of galaxies and super
        clusters are filled out with 90 percent of dark matter [26]. Rubin et al [27] have
        found that the rotation curve profile of a spiral galaxy is flat outside a central galactic
        region and they have suggested the energy density varies as $\frac{1}{r^2}$ of non luminosity
        matter ( dark matter ) which should contribute to the flatness of the rotation curves [28].
        Nucamendi and others [29-31] have suggested that the monopole ( its energy density proportional
        to $\frac{1}{r^2}$) could be the galactic dark matter in the spiral galaxies.
        In a recent work, Lee and Lee [32] studied global monopole field ( within weak field approximations )
        as a candidate for galactic dark matter in the scalar tensor theory of gravity inspired by
        the work of Nucamendi. The measurements, made by Persic et al [33] of rotation
        curves in spiral galaxies show that the coplanar orbital motion of gas in the outer parts of these
        galaxies keeps more or less a constant value up to several luminous radii.

        The renewed interest in the scalar theory of gravitation is mainly due to the belief that, at least
         at sufficient high energy scale, gravity becomes
scalar tensorial in nature and
        these theories are important in the very early Universe [34-35].
        In this article, we investigate the space-time of a global monopole field as a candidate
        for galactic dark matter in the context of scalar tensor theory.
        The gravitational field equations for a global monopole in scalar tensor theory written
        in Dicke's revised units [36], in general
        \begin{equation}
                G_{ab} = -T_{ab}-
                \frac{1}{2}\phi^{-2}(2\omega+3)[
                \phi_{,a}\phi_{,b} -  g_{ab} \phi_{,\alpha}\phi^{,\alpha}
                ]
            \label{Eq1}
          \end{equation}

        where $T_{ab}$ is the energy momentum tensor due to the monopole field, $\phi$
        is the Brans-Dicke
        scalar
        field and $\omega$ is
         the Brans-Dicke parameter.
        The energy momentum tensor of a static global monopole can be approximated ( outside the core) as [25]

        \begin{equation}
T_t^t  =  T_r^r  = \frac{\eta ^2}{r^2} ;   T_\theta^\theta  =
T_\phi^\phi =  0 .
         \label{Eq2}
          \end{equation}

        where
$\eta$ is the energy scale of symmetry breaking . As a result of
spherically symmetry, we consider $\phi=\phi(r)$ and line element
\begin{equation}
               ds^2=e^\nu dt^2-e^\beta dr^2-r^2d\Omega_2^2
         \label{Eq3}
          \end{equation}

Substituting this in equation (1) and taking into account
equation(2), we obtain the following set of equations
\begin{equation}
  e^{-\beta}[\nu^{\prime\prime} +
  \frac{1}{2}(\nu^\prime)^2-\frac{1}{2}\nu^\prime \beta^\prime +  \frac{\nu^\prime -
  \beta^\prime}{r}] = -\frac{k}{2} (\psi^\prime)^2 e^{-\beta}
         \label{Eq4}
          \end{equation}
\begin{equation}
  e^{-\beta}[\frac{1}{r^2} +\frac{\nu^\prime}{r}]-\frac{1}{r^2}
  = - \frac{\eta^2}{r^2}+\frac{k}{2} (\psi^\prime)^2 e^{-\beta}
         \label{Eq13}
          \end{equation}
\begin{equation}
 e^{-\beta}[\frac{1}{r^2}-\frac{\beta^\prime}{r}]-\frac{1}{r^2}
 = - \frac{\eta^2}{r^2}-\frac{k}{2} (\psi^\prime)^2 e^{-\beta}
        \label{Eq14}
         \end{equation}
\cen{[ where $\psi = ln\phi$ and $k =   \frac{1}{2} ( 2\omega +
3)$ and '$^\prime$' refers to differentiation  with respect to
radial
coordinate ]} \\

The wave equation for the Brans-Dicke scalar field is given by
\begin{equation}
              \psi^{\prime\prime}
              +\frac{2}{r}\psi^\prime+\psi^\prime[\frac{\nu^\prime}{2}-\frac{\beta^\prime}{2}]
                            =-\frac{ e^{\beta}{\eta^2}}{kr^2}
           \label{Eq10 }
     \end{equation}

Matos et al [28-29] have examined the possibility and type of dark
matter that determining the geometry of a space-time where the
flat rotational curves could be explained.

We are to consider global monopole in scalar tensor theory as a
candidate for galactic dark matter. To discuss rotational curves
in the galaxy, we consider the circular motion of stars in
space-time with metric (3). Now the Lagrangian for a test
particle traveling on this space-time described by (3) is
\begin{equation}
              2L = e^\nu (\dot{t})^2-e^\beta (\dot{r})^2-r^2(\dot{\theta})^2-r^2 sin^2\theta (\dot{\phi})^2
           \label{Eq10 }
     \end{equation}

    \cen{[ where  dot refers to differentiation  with respect to proper time]} \
          We consider the case, $ \theta = \frac{\pi}{2}$.
     We get generalized momenta from (8) as
     \begin{equation}
              p_t = E = e^\nu \dot{t}
           \label{Eq10 }
     \end{equation}
          \begin{equation}
              p_r = -e^\beta \dot{r}
           \label{Eq10 }
     \end{equation} \begin{equation}
              p_\phi = -W = r^2\dot{\phi}
           \label{Eq10 }
     \end{equation}
     where E is the total energy and W is angular momentum.
     Using (8),(9) and (11), the geodesic equations read
     \begin{equation}
             (\dot{r})^2 + V_{eff} (r) = 0
           \label{Eq10 }
     \end{equation}
     where $  V_{eff} (r) =    e^{-\beta}[ 1 + \frac{W^2}{r^2}- E^2e^{-\nu}] $.
     We require the following conditions for stars to have circular motions [37]

            \begin{equation}\dot{r}=0,   \frac{\partial{V_{eff}}}{\partial {r}} = 0,
             \frac{\partial^2{V_{eff}}}{\partial {r}^2}>0
            \label{Eq10 }
     \end{equation}
      Following [37], it is found that the tangential velocity of the test particle is
           \begin{equation}
             v^{tangential} = v^\phi = \surd  [\frac{r( e^\nu)^\prime}{2e^\nu}]
           \label{Eq10 }
     \end{equation}
    \cen{[ where '$^\prime$' refers to differentiation  with respect to radial
coordinate ]} \\
   It is easy to show that if flat rotation curves are required, it arises the following flat
   curve condition from (14) i.e.
   \begin{equation}
             e^\nu = B_0r^l ,
   \label{Eq10 }
     \end{equation}
    where $l$ is given by        $l = 2(v^\phi)^2$  and $B_0$ is an integration constant.
   The observed rotational curve profile in the dark matter dominant region is such that the rotational
   velocity $v^\phi$ more or less a constant. For a typical galaxy
   the rotational velocities are $v^\phi \sim 10^{-3} ( 300 km / s
   ) $ [29].
   \begin{figure}[htbp]
    \centering
        \includegraphics[scale=.8]{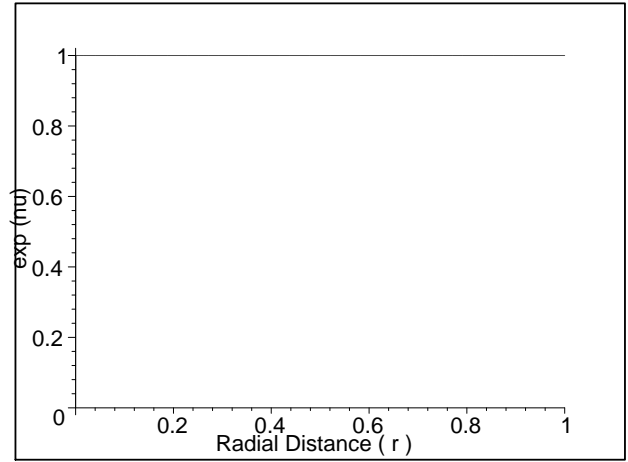}
        \caption{The diagram for $e^\nu$ ( taking $B_0 = 1$ and
        $v^\phi \sim 10^{-3}$
        )}
   \label{fig:shape1}
\end{figure}

\pagebreak

   By (6) -(5) -2(4), one gets,
   \begin{equation}
  \nu^{\prime\prime} +
  \frac{(\nu^\prime)^2}{2} +\frac{2\nu^\prime}{r} = \frac{\beta^\prime\nu^\prime}{2}
         \label{Eq4}
          \end{equation}
   This implies
   \begin{equation}
  e^\nu(\nu^\prime)^2 =  \frac{e^\beta b^2 }{r^4}
         \label{Eq4}
          \end{equation}
   where b is an integration constant.
   From (17), by taking into account (15), we get
   \begin{equation}
  e^\beta  =  \frac{B_0l^2 r^{l+2}}{b^2}
         \label{Eq4}
          \end{equation}

 \begin{figure}[htbp]
    \centering
        \includegraphics[scale=.8]{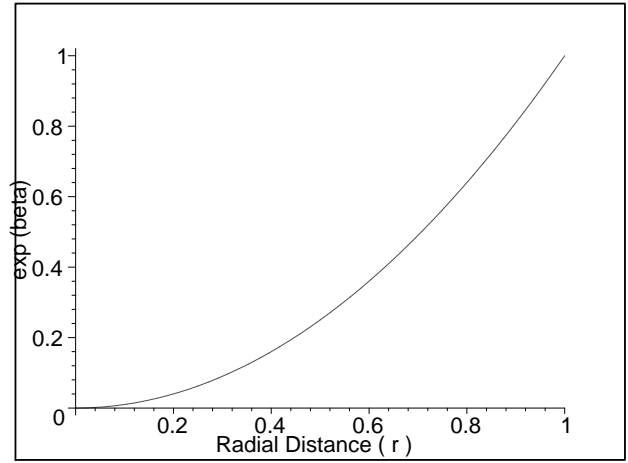}
        \caption{The diagram for $e^\beta $ ( taking $B_0l^2 = b^2$ and
        $v^\phi \sim 10^{-3}$
        )}
   \label{fig:shape1}
\end{figure}

\pagebreak

   Also from wave equation (7), by using (15) and (18), we get
    \begin{equation}
  \psi  =  -  \frac{B_0l^2 r^{l+2}\eta^2}{kb^2( l+2)^2} + D ln r + F
         \label{Eq4}
          \end{equation}
          where D and F are integration constants.
   One should note that actual Brans-Dicke scalar field
    \begin{equation}\phi = e^\psi  =  r^D exp [ F- \frac{B_0l^2 r^{l+2}\eta^2}{kb^2( l+2)^2} ]
         \label{Eq4}
          \end{equation}
   If we take D and F to be zero, then
   \begin{equation}\phi =  exp [ -  \frac{B_0l^2 r^{l+2}\eta^2}{kb^2( l+2)^2} ]
         \label{Eq4}
          \end{equation}
   Here one can see that if $\omega\rightarrow\infty, \phi\rightarrow 1 $ and $\phi\rightarrow 0$
    as $r\rightarrow\infty$.
\begin{figure}[htbp]
    \centering
        \includegraphics[scale=.8]{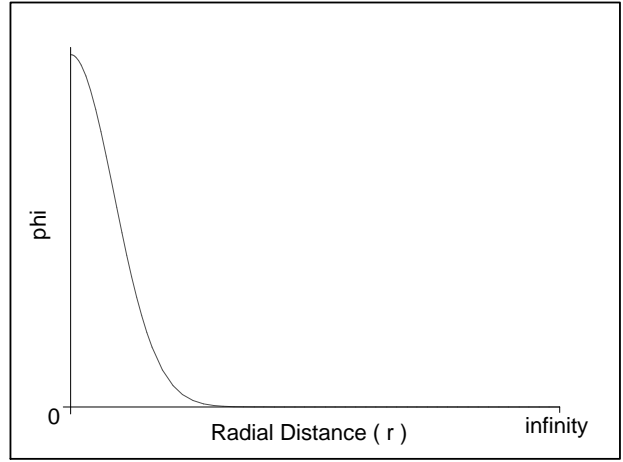}
        \caption{The diagram for Brans-Dicke scalar field $\phi $  ( taking $B_0l^2\eta^2= kb^2
        (l+2)^2 $ and
        $v^\phi \sim 10^{-3}$
        )}
   \label{fig:shape1}
\end{figure}

   With the flat curve condition, metric (3) becomes
   \begin{equation}
               ds^2=B_0r^l dt^2-\frac{B_0l^2 r^{l+2}}{b^2}  dr^2-r^2d\Omega_2^2
         \label{Eq3}
          \end{equation}
   This metric is analogous to the metric of general relativistic version of a matter distribution
   , where moving test particles follow a constant rotational curve.
   If we choose $ R = \frac{lr^2}{2b}$, then our metric takes the form ( except for a conformal factor)
    \begin{equation}
               ds^2=dt^2-dR^2-\frac{1}{B_0} [\frac{2bR}{l}]^ {\frac{2-l}{2}}d\Omega_2^2
         \label{Eq3}
          \end{equation}

          \pagebreak

   From eq.(23), we see that the curved space-time of global monopole field as a candidate for galactic dark matter
   presents a deficit solid angle in the hypersurfaces $ t=constant$. The area of a sphere of radius R
   in this space would different from ${4 \pi} {R^2}$. Here the angular deficit is dependent on the
rotational velocity of the galaxies.
   Galaxies are composed by almost 90 percent of dark matter distributed at the halos and for a typical
   galaxy, the rotational velocity more or less a constant suggests the existence of dark matter whose
   energy density varies as $\frac{1}{r^2}$. The energy density of global monopole proportional to
    $\frac{1}{r^2}$,   so according to Nucamendi et al the monopole could be the galactic matter in the
    spiral galaxies. Assuming energy density proportional to
    $\frac{1}{r^2}$, we have found a spherically symmetric metric (22) with flat curve condition.
    It is important to note that our metric is not conformally flat and hence it represents a monopole.
    Pando, Valls and Gaboud [38] have proposed that topological defects are responsible for the structure
    formation of the galaxies. Our result is in agreement with Nucamendi that monopole could be the
    galactic dark matter in the spiral galaxies. Thus it seems monopoles ( one of the topological defects )
    take part an important contribution of the galaxy formations and  give
    indirect support of Pando et al proposal.

        { \bf Acknowledgements }

           F.R is thankful to Jadavpur University and DST , Government of India for providing
          financial support under Potential Excellence and Young
          Scientist scheme . MK has been partially supported by
          UGC,
          Government of India under Minor Research Project scheme. We are grateful to the anonymous referee for his
valuable comments and constructive suggestions. \\

          \pagebreak

%\begin{figure}[p]
%\includegraphics*[450,350]{fig1.bmp}
%\caption{Variation of deflection of the circular plate}
%\end{figure}

\end{document}